\documentclass[aps,prl,reprint]{revtex4-1}

\usepackage{graphicx}
\usepackage{placeins}
\usepackage{amsmath,amssymb}
\usepackage[colorlinks=true]{hyperref}		
\usepackage{subfigure}
\usepackage{natbib}

\newcommand\ea{{\em et al.}}

\newcommand\degC{$^\circ$C}

\begin{document}

\title{Observation of electromagnetically induced transparency in evanescent fields}
\author{R.~Thomas$^1$}
\author{C.~Kupchak$^1$}
\author{G.~S.~Agarwal$^2$}
\author{A.~I.~Lvovsky$^{1,3}$}
\affiliation{$^1$Institute for Quantum Information Science, University of Calgary, Calgary, Alberta T2N 1N4, Canada}
\affiliation{$^2$Department of Physics, Oklahoma State University, Stillwater, Oklahoma 74078-3072, USA}
\affiliation{$^3$Russian Quantum Center, 100 Novaya St., Skolkovo, 
Moscow region, 143025, Russia}

\begin{abstract}

We observe and investigate, both experimentally and theoretically, electromagnetically-induced transparency experienced by evanescent fields arising due to total internal reflection from an interface of glass and hot rubidium vapor. This phenomenon manifests itself as a non-Lorentzian peak in the reflectivity spectrum, which features a sharp cusp with a sub-natural width of about 1 MHz. The width of the peak is independent of the thickness of the interaction region, which indicates that the main source of decoherence is likely due to collisions with the cell walls rather than diffusion of atoms. With the inclusion of a coherence-preserving wall coating, this system could be used as an ultra-compact frequency reference.

\end{abstract}

\maketitle


Electromagnetically induced transparency (EIT) has been studied in many different quantum systems such as atomic vapors \cite{Boller1991}, superconducting \cite{Baur2009,Sillanpaa2009} and optomechanical architectures \cite{Agarwal2010,Teufel2011,Chan2011}.  Slowdown and storage of light pulses using EIT has been used for optical quantum memories with potential applications in  long-distance quantum communication \cite{Lvovsky2009}. EIT is also a promising system for the implementation of giant optical nonlinearities, which will permit deterministic quantum optical computing \cite{Bajcsy2009,Peyronel2012}.

While most fundamental EIT studies were done with free-space optical fields, practical applications of this phenomenon require guided fields. This is particularly important for achieving high optical nonlinearities, because guided optical fields can interact with EIT media over extended lengths . Guided fields also eliminate spatial effects in these interactions, thereby increasing quantum optical gate fidelity \cite{He2011}. Particularly promising in this context are optical fibers of submicron diameter, which, when embedded into an atomic gas, allow strong coupling between the light and atoms via evanescent fields \cite{LeKien2009,Spillane2008}.

EIT has also been used as an atomic frequency standard \cite{Vanier2005}, as the EIT linewidth can be many orders of magnitude smaller than the natural absorption linewidth of typical atomic transitions.  Transmission linewidths on the order of 100 Hz have been achieved using polymer coated vapour cells \cite{Klein2006}, and optical clocks based on non-polymer coated cells have been constructed \cite{Mikhailov2010}. Achieving similar precision in \emph{microscopic} cells will allow compact frequency standards, thereby dramatically enhancing the precision of portable geopositioning systems. Because evanescent fields have penetration depths on a scale of single microns, they offer a favorable venue for developing such standards.

The above examples show the importance of EIT in evanescent fields in both fundamental and applied aspects of quantum technology. However, to date there existed no conclusive experimental evidence of this phenomenon. The present paper accomplishes this result and provides its detailed theoretical and experimental study.

\begin{figure}[h]
\centering
\includegraphics[width=2.5in]{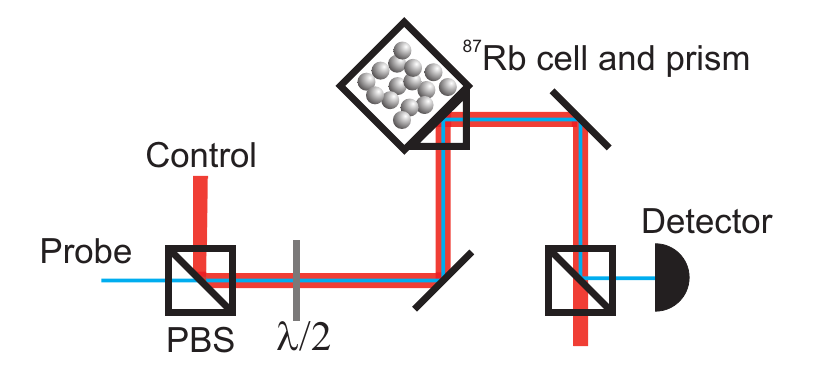}
\caption{Schematic of the experimental setup used to observe EIT with evanescent fields.   PBS: polarizing beam splitter, $\lambda/2$: half-wave plate.}
\label{fg:setup}
\end{figure}
We observe EIT with control and signal beams totally reflected from an interface between glass and hot rubidium vapor in a macroscopic cell.  Both fields are evanescent inside the vapor.  The fact that we can achieve EIT in this configuration is remarkable, as the size of the interaction region is on the order of the wavelength of the optical field.  One might expect that transit broadening and collisions with the cell wall would preclude observable ground-state coherence.  However, we observe a pronounced peak in the reflectivity spectrum which is characteristic of EIT.

Our work is distinct from experiments with nanocells \cite{Sargsyan2011,Pashayan-Leroy2007,Lenci2009}.  Although in both cases one of the dimensions in the interaction volume is microscopic, in nanocells the atoms are confined in that dimension.  In our case, the atoms can travel freely in the dimension perpendicular to the interface.

A related phenomenon has been theoretically investigated by Harris \cite{Harris1996}. In that study, the control field propagating through a plasma induced the transparency, which changed the probe field's character from evanescent to traveling. In our case, the probe field remains evanescent independently of the control field intensity.

We detect EIT by means of selective reflection \cite{Guo1996,Wang1997,Gross1997,Akulshin1989}. This is a  spectroscopic technique in which one obtains information about the susceptibility of a substance by measuring the reflectivity of its surface which is related to the susceptibility via the Fresnel equations. Typically, selective reflection experiments are done for incidence angles near to normal, which provides information about the real part of the susceptibility.  We, on the other hand, work in the regime of total reflection, and then the reflectivity mimics the imaginary part of the susceptibility.  In this way, anomalies in the absorption spectrum become manifest.

We employ the experimental setup shown in Fig.~\ref{fg:setup} to observe the effect of EIT on selective reflection.
Our EIT medium is isotopically pure rubidium-87 vapor with 50 Torr of neon as a buffer gas contained in a glass cell.  The cell is enclosed in an oven with a single layer of $\mu$-metal shielding and kept at approximately 170\degC, although the interface extends outside the oven by about 1 cm and is therefore at a colder temperature.  We use a right-angle prism to couple light to the cell, and the prism-cell interface is filled with index-matching gel to reduce losses.  We mount the system on a New Focus 5-axis translation stage (model number 9081) which allows us to vary the angle of incidence with a stated precision of about 100 $\mu$rad.

\begin{figure}[b]
\centering
\includegraphics[width=3.25in]{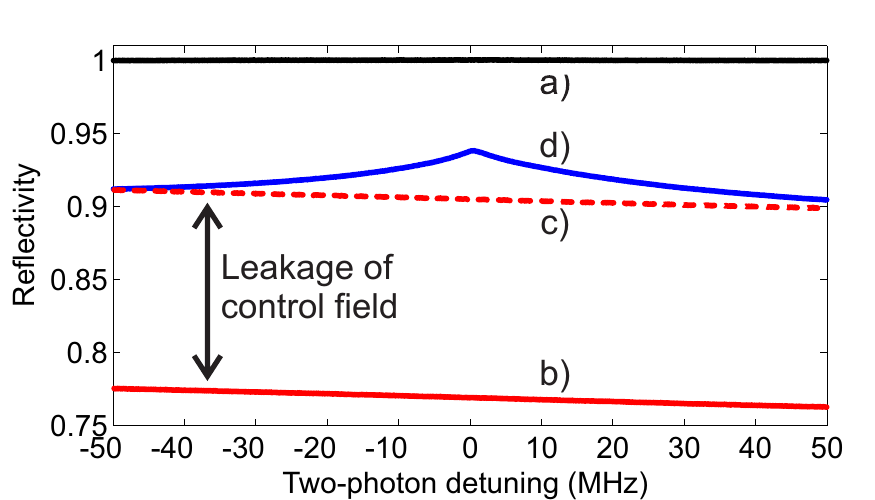}
\caption{Power of the probe field reflected from the interface under various conditions.  a) Probe field scanning far off-resonance, control field present.  b) Probe field scanning over resonance, control field absent.  c) Same as b), but the control field power that leaks through the filtering PBS is added to show consistency between (b) and (d).  d) Probe field scanning over resonance, control field present, resulting in an EIT window.  Control field power is 75 mW, and $\theta-\theta_c\approx 3$ mrad.}
\label{fg:Contrast}
\end{figure}

We use a Coherent MBR-110 Ti:Sapphire laser as the source of our control field and a home-built external-cavity diode laser to generate the probe field.  The probe field is tuned to the $|F=1,m_F\rangle\rightarrow|F'=1,m_F\rangle$ transition of the $^{87}$Rb D1 line, and the control field is tuned to the $|F=2,m_F\rangle\rightarrow|F'=1,m_F\rangle$ transition.  We stabilize the frequency difference between the two fields using a home-built phase lock \cite{Appel2009}.  The probe field is polarized perpendicularly to the plane of incidence and the control field is polarized in the plane, so we can separate the two fields using a single polarizing beam-splitter.

We probe the system by measuring the reflected power of the probe field as a function of the two-photon detuning at various values of the incidence angle $\theta$ near the critical angle $\theta_c=41.8^\circ$. An example of the reflectivity spectrum for a specific $\theta>\theta_c$ is shown in Fig.~\ref{fg:Contrast}.
We observe a peak in the reflectivity when the frequency of the probe field is scanned over the two-photon resonance and the control field is present. From curves (a) and (d) in Fig.~\ref{fg:Contrast}, we measure the contrast of our EIT line to be 35\%.

\begin{figure}[b]
\centering
\includegraphics[width=3.5in]{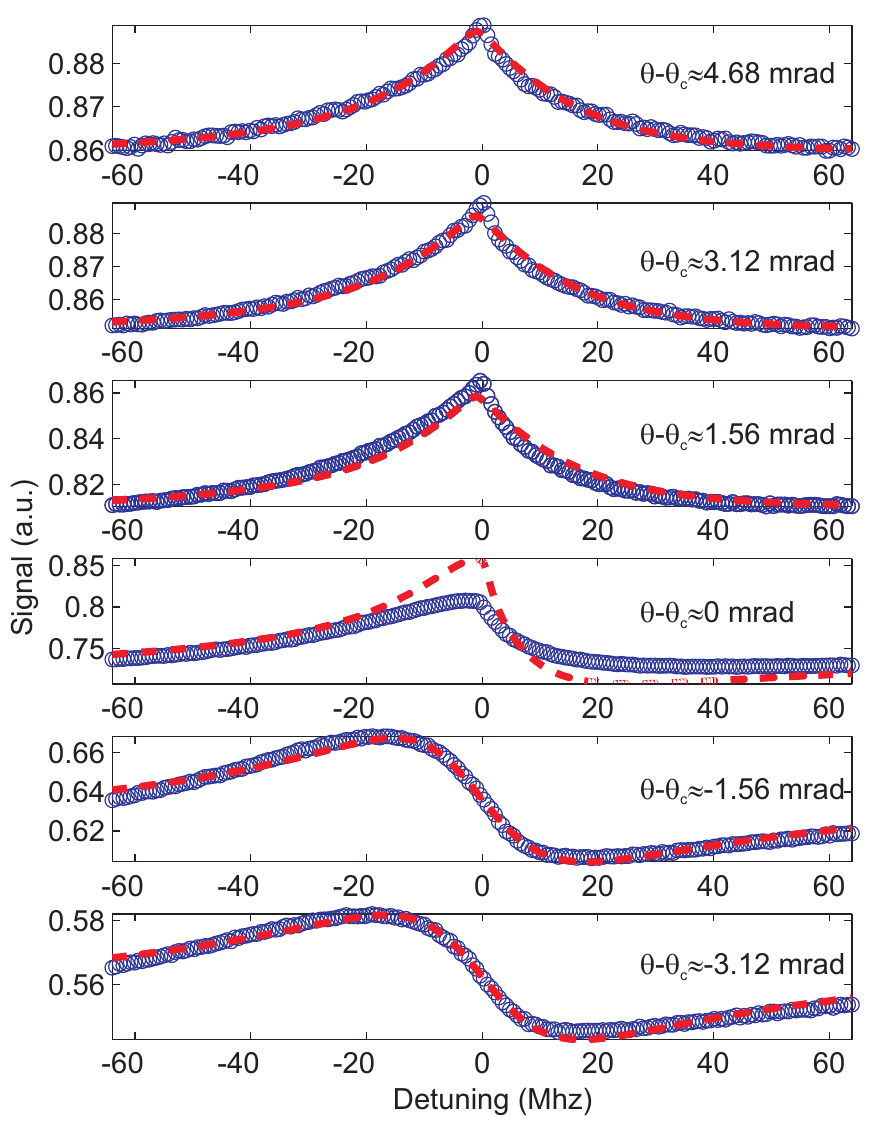}
\caption{Measured reflected probe field power with the control field present (blue circles) and theoretical fits (red dashed lines) for different values of the incidence angle. The control field power is the same for all plots.}
\label{fg:Data1D}
\end{figure}

The reflected power as a function of the two-photon detuning for several incidence angles is shown in Fig.~\ref{fg:Data1D}.
One can identify two different regimes. For $\theta<\theta_c$, the reflection line shape is dispersive in appearance, and when $\theta>\theta_c$ the line shape resembles a traditional transparency window with a sharp, cusp-like feature at the two-photon resonance.  In the direct neighborhood of the critical angle, the line shape is a hybrid between the two different regimes.

We now briefly describe a simple, analytical model for the reflection spectra based on a general theory of selective reflection by Nienhuis \ea~\cite{Nienhuis1988}.  We consider two plane waves, corresponding to the probe and control fields, incident on an interface in the $x-y$ plane.  The incident, reflected and transmitted probe fields are denoted $E_1(z)$, $E_2(z)$, and $E_3(z)$, respectively.  In the half-space $z<0$ we have a dielectric with an index of refraction of $n_1$, and in the half-space $z>0$ we have a low density atomic vapor with index of refraction $n_2(z)=1+\chi(z)/2$ where $\chi(z)$ is the susceptibility of the vapor.  For an EIT medium in the weak probe limit we have the standard expression \cite{Scully}
\begin{equation}
\chi(z)=\frac{N d^2\sqrt{\pi}}{\epsilon_0\hbar}\frac{\delta+i\gamma}{\Omega_c^2|e^{2i\beta k_0 z}|-(\delta+i\gamma)(\delta+\Delta_c+i\Gamma/2)}
\label{eq:EITSusc}
\end{equation}
where $d$ is the dipole matrix element of the atomic transition, $N$ is the atomic number density, $\delta$ is the two-photon detuning, $\gamma$ is the ground-state dephasing, $\Delta_c$ is the detuning of the control field from resonance, $\Omega_c$ is the control field Rabi frequency in the atomic vapor at the interface, $\Gamma$ is the sum of radiative, collisional and Doppler widths \cite{Figueroa2006}, $k_0$ is the wave number of the control field, and $\beta=\sqrt{1-n_1^2\sin^2\theta}$. Dependent on whether $\beta$ is imaginary ($\theta>\theta_c=\sin^{-1}(1/n_1)$) or real ($\theta\leq\theta_c$), the susceptibility does or does not depend on $z$ through the exponential in the denominator.

When the susceptibility is independent of $z$ we can immediately use the Fresnel reflection coefficient for S-polarization \cite{Jackson}
\begin{equation}
E_2(0)=\frac{\cos\theta-\sqrt{(\frac{n_2}{n_1})^2-\sin^2\theta}} {\cos\theta+\sqrt{(\frac{n_2}{n_1})^2-\sin^2\theta}}E_1(0)\equiv r(\chi)E_1(0)
\label{eq:Fresnel reflection}
\end{equation}
 to determine the spectrum of the reflected field.  The dispersive shape of the EIT line in this regime is due to the sharp dependence of the reflectivity on the real part of the susceptibility.  

For $\theta>\theta_c$, the control and probe fields are evanescent and decay exponentially into the vapour.  In this case $\beta$ is imaginary, and we let $\beta=i\eta$.  We assume that $\chi/\eta\ll 1$ and use a Born series to determine the corrections to the reflected field due to the presence of the atomic vapour.  We write the reflected and transmitted fields as sums of terms $E_{2,3}(z)=\sum_{n}E_{2,3}^{(n)}(z)$ which are determined from a reduced form of the wave equation
\begin{equation}
	\begin{array}{lr}
		(\partial_z^2-k_0^2\eta^2)E_3^{(n)}(z)=-k_0^2\chi(z)E_3^{(n-1)}(z)& z>0\\
		(\partial_z^2+n_1^2 k_0^2\cos^2\theta)E_2^{(n)}(z)=0& z<0
	\end{array}
\label{eq:WaveEq}
\end{equation}
and the relevant boundary conditions from Maxwell's equations.  We use a pair of Green's functions which satisfy the boundary conditions \cite{Tai} to solve Eq.~\eqref{eq:WaveEq}, and to first order ($n=1$) the amplitude of the reflected field is
\begin{align}
E_2(0)&=r(\chi=0) \left(1+\frac{2i n_1k_0\cos\theta}{\eta(n_1^2-1)}\int_0^{\infty}\chi(z)e^{-2k_0\eta z}dz\right)\nonumber\\
&=r(\chi=0)\left(1+\frac{i n_1\cos\theta}{\eta(n_1^2-1)}\chi_{\textrm{eff}}\right)
\label{eq:ReflectCorr}
\end{align}
where
\begin{equation}
\chi_{\textrm{eff}}=\frac{N d^2\sqrt{\pi}}{\epsilon_0\hbar}\frac{\delta+i\gamma}{\Omega_c^2}\ln\left(1-\frac{\Omega_c^2}{(\delta+i\gamma)(\delta+\Delta_c+i\Gamma/2)}\right)
\label{eq:Susc2}
\end{equation}
is the effective susceptibility.  The latter is the susceptibility that a homogeneous medium would require in order to have the same reflectivity as the medium with spatially dependent susceptibility studied here.  The reflectivity $R=|r|^2$ is then linear in the imaginary part of $\chi_{\textrm{eff}}$.

The cusp feature in the EIT spectra is due to the non-trivial dependence of the polarization on penetration depth and is reflected by the logarithmic term in Eq.~\eqref{eq:Susc2}.  One can see its emergence by looking at the imaginary part of Eq.~\eqref{eq:Susc2} in the limit of $\Gamma\gg\delta$ and $\gamma=0$.  Its derivative behaves as $\tan^{-1}\left(\frac{2\Omega_c^2}{\Gamma\delta}\right)$ which has a discontinuity at $\delta=0$.

A more intuitive explanation of the cusp is that since both the probe and the control fields are evanescent, atoms near the interface will be exposed to a much stronger control field than atoms that are further away from the interface.   Since the full-width at half-maximum (FWHM) of the standard transparency window for homogeneous plane waves is given by $2\gamma+\frac{4\Omega_c^2}{\Gamma}$  \cite{Figueroa2006}, atoms that are close to the interface will exhibit a wider transparency window than those that are more distant.  The most distant atoms will exhibit a linewidth that is dictated solely by the ground state dephasing of the system. In addition, atoms that are close to the interface will contribute more to the susceptibility than those that are farther away due to the limited penetration of the probe field into the vapor.  Similar mechanisms for producing non-Lorentzian lineshapes have been explored by Ta\ifmmode \breve{\imath}\else \u{\i}\fi{}chenachev {\it et al.} \cite{Tachenachev2004} as well as  Le Kien and Hakuta \cite{LeKien2009}.

We fit our model to the acquired reflectivity spectra as shown in Fig.~\ref{fg:Data1D} by dashed lines. For all curves, we fix $\Delta_c=2\pi\times 50$ MHz and $\gamma=2\pi\times 0.5$ MHz. We fit for $\Omega_c$ and the number density in the two regimes $\theta<\theta_c$ and $\theta>\theta_c$ separately. The number density changes from $6.8\times 10^{18}$ m$^{-3}$ for $\theta<\theta_c$ to $6.0\times 10^{18}$ m$^{-3}$ for $\theta>\theta_c$.  The value of the control field Rabi frequency changes from $2\pi\times 160$ MHz for $\theta<\theta_c$ to $2\pi\times 210$ MHz for $\theta>\theta_c$, indicating that when the fields become evanescent the transmission peak becomes broader. We also allow the reflectivity baseline of each curve to vary freely. This is justified because both the inner and outer cell windows have antireflection coating; as a result, the angle-dependent reflectivity of the cell window in the absence of the atomic vapor is difficult to predict theoretically.

The theoretical fits show satisfactory agreement with the experimental data except for angles very close to critical. The latter discrepancy is due to the dependence of the critical angle itself on the frequency of the probe field. For $\theta\approx\theta_c$, tuning the probe field switches back and forth between the two lineshape regimes. Furthermore, the assumption of $\chi/\eta\ll 1$ becomes invalid for $\theta$ in the direct proximity of $\theta_c$, resulting in the breakdown of the Born approximation.  

\begin{figure}[h]
\centering
\includegraphics[width=3.25 in]{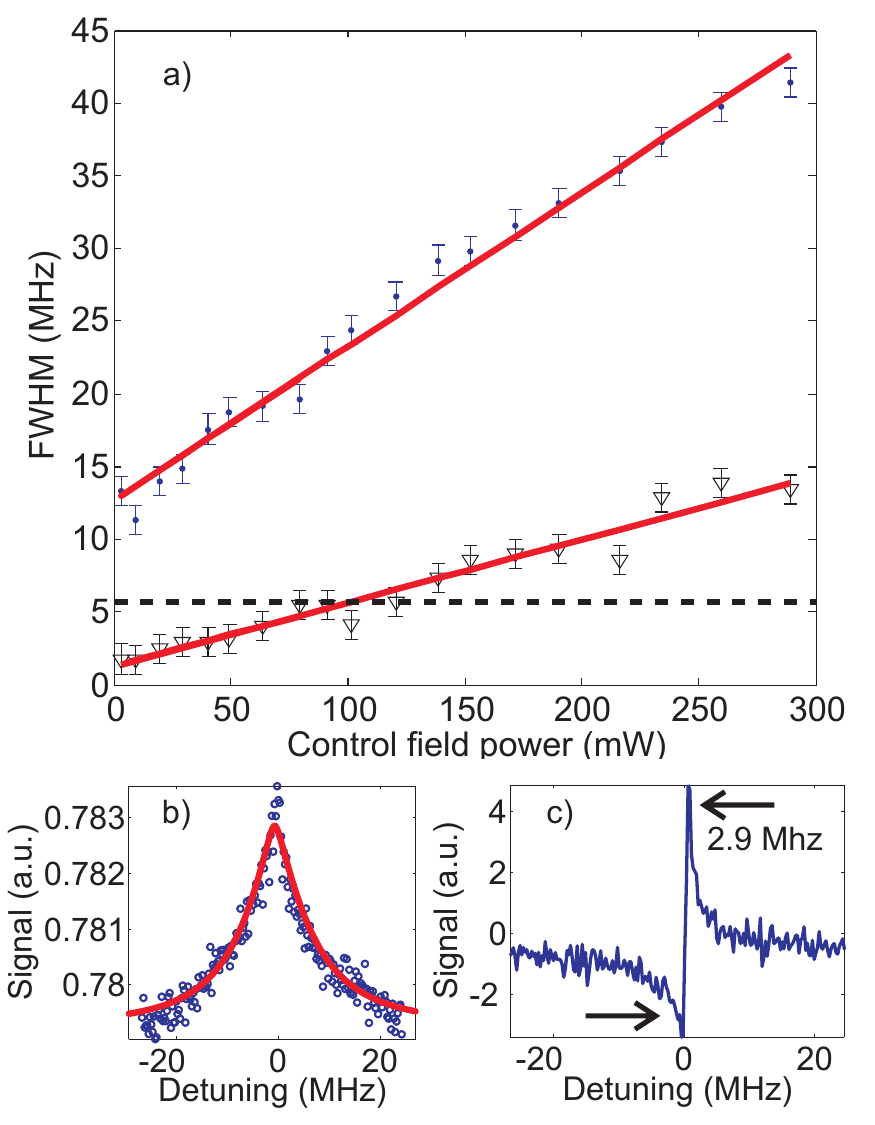}
\caption{(a) Width of the evanescent EIT line as a function of the control field power, showing the measurements for both the pedestal (blue circles) and the cusp (black triangles). Solid lines are linear fits, and the dashed line indicates the width of the natural rubidium absorption line (5.77 MHz \cite{Steck}).  
(b) Reflection spectrum and (c) its derivative with a measured cusp FWHM of 1.77 MHz.  The solid line in (b) is a fit to Eq.~\eqref{eq:ReflectCorr}; note that the experimentally observed cusp is narrower than predicted theoretically. The data in (b) and (c) correspond to a control field power of 30 mW.}
\label{fg:Width}
\end{figure}

Of central interest to us is the shape and width of the EIT line, particularly those of the cusp. These will likely be important in any experiments that use evanescent EIT for applications in quantum information or metrology.  We performe two additional experiments to that effect, measuring the EIT linewidth as a function of the control field power and the incidence angle. In both experiments, we measure the width of the cusp as well as that of the entire EIT feature, which we call the ``pedestal''. The FWHM of the pedestal is evaluated directly from the reflection spectra. The width of the cusp is estimated from the frequency difference between the maximum and minimum of the derivative of the reflectivity spectrum [Fig.~\ref{fg:Width}(c)] measured with a differentiating circuit. This difference is multiplied by $\sqrt{3}$ to give an effective FWHM.

The dependencies of the linewidths on the control field power are shown in Fig.~\ref{fg:Width}(a). We find that for a significant range of control field powers the width of the cusp is less than the natural linewidth of the D1 transition in rubidium.  This is strong evidence that the observed reflectivity peak is due to ground state coherence rather than non-linearities associated with, e.g., optical pumping.

As expected, both widths show increase with the control field power. However, the distinct linear dependence observed does not agree with our model, which predicts that the linewidth of the cusp should be related to the square root of the control field power, and that at zero control field power the width of the pedestal should be equal to the width of the cusp. These discrepancies will be the subject of further study. In particular, the shape of the pedestal can be affected by velocity changing collisions of the rubidium atoms with the cell window \cite{Berman:86}. On the other hand, the observed additional narrowing of the cusp [Fig.~\ref{fg:Width}(b)] may be attributable to diffusion-induced Ramsey-narrowing \cite{Xiao2006}.


\begin{figure}[h]
\centering
\includegraphics[width=3.25 in]{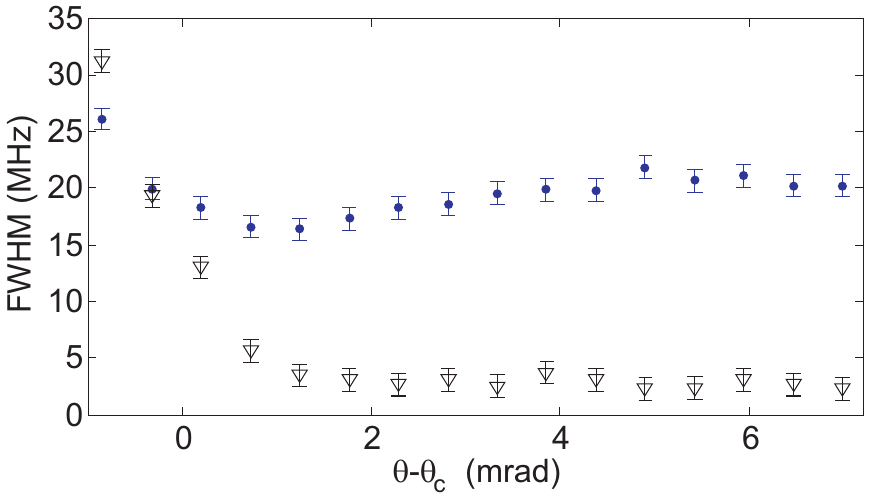}
\caption{FWHM of both the pedestal (blue circles) and cusp (black triangles) as a function of $\theta-\theta_c$.  Constant width for $\theta>\theta_c$ shows that the primary decoherence mechanism is collisions with the interface.  Data taken for a control field power of approximately 70 mW.}
\label{fg:Width2}
\end{figure}
We also characterize the widths of the pedestal and the cusp as a function of angle.  The results are shown in Fig.~\ref{fg:Width2}. For $\theta\gg\theta_c$ we can see that the widths of both the cusp and the pedestal are largely insensitive to the incidence angle. If diffusion of atoms contributed significantly to the width of the transmission peak, we would expect to see an increase in the pedestal width as the incidence angle increases. This indicates that diffusion is not a significant effect in our system.  Furthermore, the expected Rb-Rb spin-exchanging collision rate at our density of $6.0\times 10^{18}$ m$^{-3}$ is $2\pi\times 1$ kHz \cite{Moos1964} which is much less than the measured decoherence rate. We conclude that the primary cause of decoherence is due to interface collisions, and such decoherence can be substantially reduced by using polymer coated cell windows \cite{Balabas2010}.


In summary, we showed that it is possible to obtain EIT transmission peaks with evanescent probe and control fields with central linewidths that are less than the natural linewidth of the rubidium D1 transition.  Our spectra are in reasonable agreement with a simple analytical model based on a depth-dependent susceptibility, resulting in a lineshape which is a sum of Lorentzian transmission windows of varying widths.  We demonstrated the surprising result that the linewidth of the transmission peak is independent of the angle of incidence and hence of the size of the interaction region down to a single wavelength.  With the introduction of a polymer cell window coating, the frequency stability and compactness could be comparable with or better than frequency references based on microcells \cite{Knappe2005,Lenci2009} but in a more easily manufactured system.

\bibliography{EvEITBib,EvEITBib2}

\end{document}